\newcommand{\CLASSINPUTtoptextmargin}{1in}
\documentclass[conference,10pt]{IEEEtran}

\IEEEoverridecommandlockouts
\usepackage{subcaption}
\usepackage{cite}
\usepackage{graphicx}
\usepackage{graphics}
\usepackage[cmex10]{amsmath}
\usepackage{epsfig,epic}
\usepackage{array}
\usepackage{epstopdf}
\usepackage{fancyhdr}

\usepackage{multirow}
\usepackage{multicol}
\usepackage{ctable}
\usepackage{amsfonts}
\usepackage{comment}

\newcommand\given[1][]{\:#1\vert\:}

\def\argmax{\mathop{\mathrm{argmax}}}

\begin{document}
\title{Reliable and Low-Complexity MIMO Detector Selection using Neural Network}
	\author{\IEEEauthorblockN{Shailesh Chaudhari, HyukJoon Kwon, and Kee-Bong Song}
	\IEEEauthorblockA{SoC R\&D, Samsung Semiconductor Inc.,
		San Diego, USA\\
		Email: \{shailesh.dc, hyukjoon.k, keebong.s\}@samsung.com}
}

% make the title area
\maketitle
\begin{abstract}
	In this paper, we propose to dynamically select a MIMO detector using neural network for each resource element (RE) in the transport block of 5G NR/LTE communication system. The objective is to minimize the computational complexity of MIMO detection while keeping the transport block error rate (BLER) close to the BLER when dimension-reduced maximum-likelihood (DR-ML) detection is used. A detector selection problem is formulated to achieve this objective. However, since the problem is high dimensional and NP-hard, we first decompose the problem into smaller problems and train a multi-layer perceptron (MLP) network to obtain the solution. The MLP network is trained to select a low-complexity, yet reliable, detector using instantaneous channel condition in the RE. We first propose a method to generate a labeled dataset to select a low-complexity detector. Then, the MLP is trained twice using quasi-Newton method to select a reliable detector for each RE. The performance of online detector selection is evaluated in 5G NR link level simulator in terms of BLER and the complexity is quantified in terms of the number of Euclidean distance (ED) computations and the number of real additions and multiplication. Results show that the computational complexity in the MIMO detector can be reduced by $\sim$10$\times$ using the proposed method.
\end{abstract}
\IEEEpeerreviewmaketitle


%\fancyhead[LO]{{\bfseries\large{SAMSUNG}}\\ {\bfseries\large{Best paper Award 2019}}}
%\fancyfoot[LO]{\small \textsuperscript{1} \textsuperscript{2} \textsuperscript{3} SSI Modem R\&D, San Diego, USA}

% Start Contents
\vspace{-3mm}
\section{Introduction}
Next-generation cellular wireless technologies such as LTE-Advanced Pro and 5G NR adopt advanced MIMO transmission techniques in a wide frequency band to achieve the multi-Gbps speed in cellular networks. However, due to large bandwidth and high datarate requirements, the power consumption becomes a bottleneck in baseband modem. Reducing the computational complexity of the baseband modem is critical to reduce the power consumption. In this paper, we propose to reduce the computational complexity of MIMO detector in the modem while maintaining low block error rate (BLER) in the transport block in 5G NR transmission. 

The MIMO detector is used to generate log-likelihood ratios (LLRs) for bits mapped to multiple resource elements (REs) in a transport block using the OFDM-based frame structure. The LLRs are further processed by the decoder to recover the bits in transport block. In order to minimize the BLER, highly accurate LLRs must be generated by the detector. The quality of LLR depends on the instantaneous channel quality in the RE as well as the complexity of the detector used. In a typical communication receiver, the same detector is used for all REs (static utilization) without taking into account channel conditions in the RE. In such static detector utilization, a high complexity detector is required to achieve low BLER. 

Numerous works in the past have been proposed to achieve BLER close to that of ML detector while reducing the computational complexity \cite{Hochwald2003, Lee2010, Rahmati2015}. A dimension reduced ML (DR-ML) detector is presented in \cite{Lee2010} which reduces the number of constellation points visited during LLR computation. Papers \cite{Hochwald2003} and \cite{Rahmati2015} present list-detectors and initial candidate reduction (ICR) detectors, respectively, which are based on the sphere decoder idea and further reduce the detector complexity by reducing the number of constellation points in LLR computation. However, these works still consider static utilization where same detector is used for all REs. In static detector utilization, there exists a trade-off between the BLER and the detector complexity, i.e., if the complexity of the detector is reduced, then the BLER increases and vice versa. To overcome this trade-off, we propose to dynamically select the detector for each RE based on instantaneous channel condition in the RE. However, the main challenge is that this is a high-dimensional, NP-hard problem and the communication receiver needs to obtain a solution quickly with a low-complexity method. Another challenge is to obtain the solution \textit{before} actually evaluating any of the candidate detector blocks. In this paper, we propose to use a multi-layer perceptron (MLP) network to obtain solution to this problem. The main contributions of this paper are summarized below.

\begin{enumerate}
	\item High-complexity, NP-hard detector selection problem is decomposed into smaller problems which independently selects detector for each RE. In order to select the detector \textit{before} evaluating any candidate detector, we train a MLP network to select the detector using features derived from instantaneous channel in the RE, received signal and noise variance.
	\item A \textit{reliable detector selection} method is proposed to process the MLP outputs and select appropriate low-complexity, yet a reliable detector to keep the BLER low.
	\item To further reduce the complexity of detector selection at run-time (online), the MLP is re-trained to incorporate \textit{reliable detector selection} into MLP training.	
\end{enumerate}

\noindent\textit{Outline}: The system model and problem formulation are presented in Section \ref{sec:system}. The proposed method is presented in Section \ref{sec:proposed_method} including generation of labeled dataset, MLP training, and the \textit{reliable detector selection} method. Results are shown in Section \ref{sec:results} and the paper is concluded in Section \ref{sec:Conclusion}.

\textit{Notations}: Vectors are denoted by bold, lower-case letters, e.g., $\mathbf{h}$. Matrices are denoted by bold, upper case letters, e.g., $\mathbf{H}$. Transpose and Hermitian transpose are denoted by $(.)^T$ and $(.)^*$, respectively. The norm of a vector $\mathbf{h}$ is denoted by $||\mathbf{h}||$. A set of integers from $a$ to $b$ is denoted by $[a,b]$.

\section{System Model and Objective}
\label{sec:system}
Consider transmission of a transport block according to 5G NR standard \cite{3gpp2018_38211}. The bits in the transport block are mapped to $R$ resource elements (REs) using a given modulation and coding scheme (MCS) which determines modulation order and code-rate. We consider a $2\times2$ MIMO system in this paper, although the proposed method can be used with higher number of MIMO layers as well. The $2\times 1$ received signal vector in the $n$th RE is given by
\begin{align}
\mathbf{y}_n = \mathbf{H}_n \mathbf{x}_n + \mathbf{n}_n = \mathbf{h}_{n,1} x_{n,1} + \mathbf{h}_{n,2} x_{n,2} + \mathbf{n}_n,
\end{align}
where $\mathbf{H}_n \in \mathbb{C}^{2\times 2}$ is the MIMO channel matrix, $\mathbf{x}_n = [x_{n,1}, x_{n,2}]^T$ is the transmitted signal vector, and $\mathbf{n}_n \sim \mathcal{CN}(0,\sigma^2 \mathbf{I})$ is the noise vector. The transmitted symbol in MIMO layer-$t$ is $x_{n,t}$ which represents $M$ bits: $b_{n,t,m}, m\in [1,M]$. 

At the receiver, the MIMO detector generates LLRs for bits $b_{n,t,m}, t\in [1,2], m\in [1,M]$ using $\mathbf{y}_n, \mathbf{H}_n$, and $\sigma^2$. The a posteriori LLR for bit $b_{n,1,m}$ in MIMO layer-1 is defined as

{\small
\vspace{-3mm}
\begin{align}
\nonumber L_{n,1,m} =& \log \left(\frac{\Pr(b_{n,1,m} = 1 | \mathbf{y}_n)}{\Pr(b_{n,1,m} = 0| \mathbf{y}_n)} \right) 
\\ =& \min\limits_{\mathbf{x} \in \mathcal{S}_1^+} \frac{||\mathbf{y}_n - \mathbf{H}_n \mathbf{x}||^2}{\sigma^2}
-\min\limits_{\mathbf{x} \in \mathcal{S}_1^-} \frac{||\mathbf{y}_n - \mathbf{H}_n \mathbf{x}||^2}{\sigma^2}
\label{eq:LLR}
\end{align}
}

\noindent where $\mathcal{S}_1^+$ is the set of vectors of constellation symbols $[x_1, \hat{x}_2(x_1)]^T$ such that the $m$-th bit in symbol $x_1$ is 1. For a given $x_1$, $\hat{x}_2(x_1)$ satisfies the following \cite{Rahmati2015}:
\begin{align}
\hat{x}_2(x_1) = \arg \min_{x_2} ||\mathbf{y}_n -  \mathbf{h}_{n,1} x_1 - \mathbf{h}_{n,2}{x}_2||^2
\label{eq:hard_decode}
\end{align}
Similarly, $\mathcal{S}_1^-$ is the set of vectors of constellation points $[x_1, \hat{x}_2(x_1)]^T$ such that the $m$-th bit in constellation $x_1$ is 0. The expression for LLRs for bits in layer-2 can be written in a similar way by replacing indices 1 with 2 in (\ref{eq:LLR}) and (\ref{eq:hard_decode}). The number of Euclidean distance (ED) computations in (\ref{eq:LLR}) is equal to the number of candidates vectors in set $\mathcal{S} = \mathcal{S}_1^+ \cup \mathcal{S}_1^-$. We consider $D=5$ candidate detectors that utilize sets of different sizes to compute the LLRs. Let $L^{(d)}_{n,t,m}$ indicate the LLR computed by detector-$d$ for the bit $b_{n,t,m}$. The detectors are indexed in ascending order of complexity as follows:
\begin{align}
\nonumber \mathcal{Z} = \{1: \text{MMSE}, 2: \text{ICR-16}, & 3: \text{ICR-32},
\\ & 4: \text{ICR-64}, 5:\text{DR-ML}\}
\end{align}
The most complex detector is DR-ML detector \cite{Lee2010} which requires $2^M$ ED computations since it utilizes a set $\mathcal{S}$ with $2^M$ elements. ICR-16, ICR-32, and ICR-64 detectors require 16, 32, and 64 ED computations, respectively \cite{Rahmati2015}. Finally, the MMSE detector is the lowest complexity detector that computes the LLR using soft de-mapper requiring no ED computations \cite{Mao2016}.

\begin{figure*}[t]
	\centering
	\begin{subfigure}[b]{0.48\textwidth}
		\centering
		\includegraphics[width=\columnwidth]{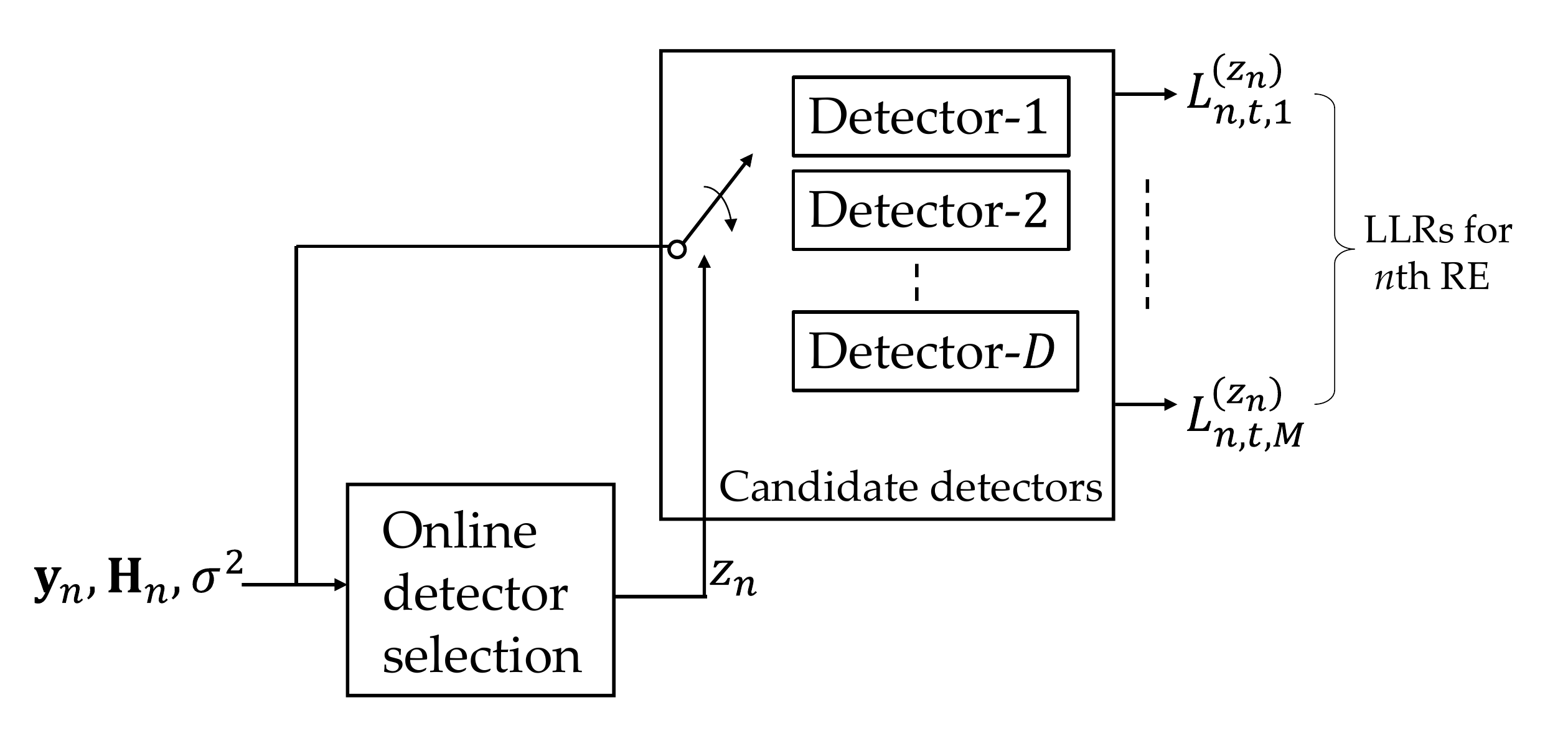}	
		\caption{{\footnotesize System model for detector selection.}}
		\label{fig:detector}
	\end{subfigure}
	\begin{subfigure}[b]{0.48\textwidth}
		\centering
		\includegraphics[width=\columnwidth]{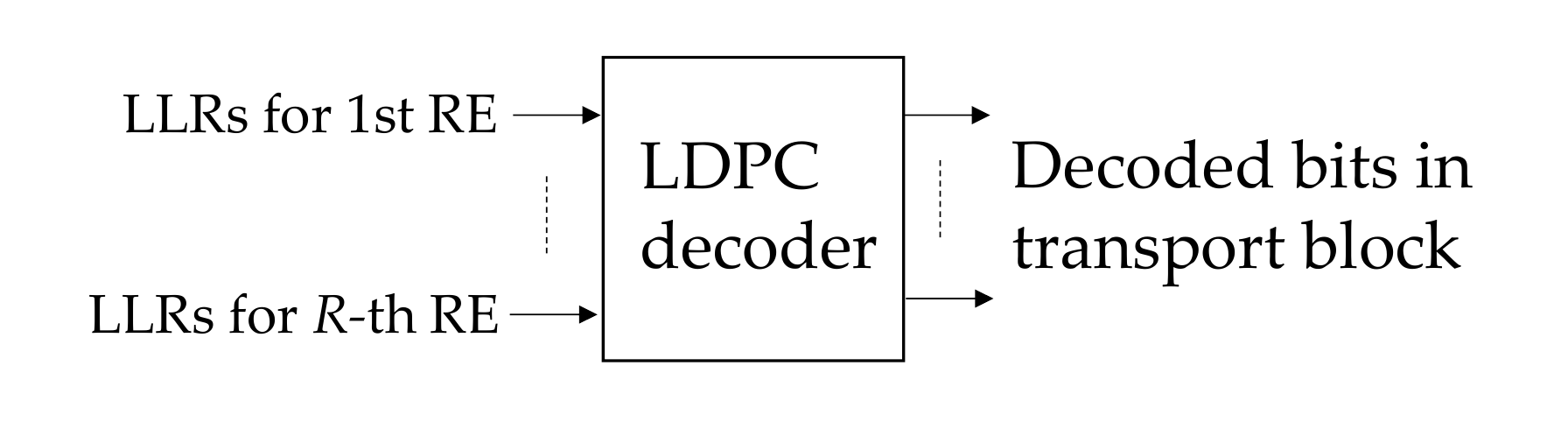}
		\caption{{\footnotesize LDPC decoding using detector LLRs.}}
		\label{fig:decoder}
	\end{subfigure}
	\caption{\footnotesize MIMO Detector and LDPC decoder at receiver.}
	\label{fig:detector_decoder}
	\vspace{-4mm}
\end{figure*}

As shown in Fig. \ref{fig:detector_decoder}, the LLRs generated by the detector are used by the LDPC decoder to decode bits in the transport block. The block error is declared whenever any one bit is in error in the decoded transport block.

\begin{comment}
The MMSE detector evaluates the LLR as follows:
\begin{align}
L^{(1)}_{n,1,m} = \min \limits_{x \in\mathcal{X}^{-}_{m}} {|\tilde{y} - \tilde{h}x|^2} - \min\limits_{x \in\mathcal{X}^{+}_{m}} {|\tilde{y}_n - \tilde{h}_n x|^2},
\label{eq:LLR_MMSE}
\end{align}
where $\mathcal{X}_m^-$ and $\mathcal{X}_m^+$ are the sets of constellation points with $m$-th bit is 0 and 1, respectively. Further, $\tilde{y}_n = \frac{\mathbf{w}^*_{n,1}\mathbf{y}_n}{\sqrt{|\mathbf{w}^*_{n,1}\mathbf{h}_{n,2}|^2 + ||\mathbf{w}_{n,1}||^2 {\sigma^2}}}$, $\tilde{h}_n = \frac{\mathbf{w}^*_{n,1}\mathbf{h}_{n,1}}{\sqrt{|\mathbf{w}^*_{n,1}\mathbf{h}_{n,2}|^2 + ||\mathbf{w}_{n,1}||^2 {\sigma^2}}}$ and $\mathbf{w}^*_{n,1}$ is the MMSE weight vector and is the  first column of $\mathbf{W}_n = \mathbf{H}_n(\mathbf{H}_n^*\mathbf{H}_n + \sigma^2 \mathbf{I})^{-1}$. The expression (\ref{eq:LLR_MMSE}) is a soft de-mapper that requires no ED computations \cite{Mao2016}. 
\end{comment}
\begin{comment}
\begin{figure}[t]
	\centering
	\includegraphics[width=\columnwidth]{Figures/system_model.pdf}	
	\caption{\footnotesize System model for online dynamic detector selection.}
	\label{fig:system_model}
	%	\vspace{-3mm}
\end{figure}
\end{comment}

\subsection{Problem formulation}
\label{sec:formulation}
The objective of this work is to select low-complexity detectors $z_n,n=1,2,...,R$ for each RE in the transport block while keeping the BLER close to the most complex DR-ML detector. Let us define $P^{DRML}_e$ as the BLER when all REs use the DR-ML detector, $P_e(\mathbf{z})$ as the BLER when the $n$-th RE uses the selected detector $z_n \in \mathcal{Z}$. Then, the objective can be mathematically described as

{\small
\vspace{-3mm}
\begin{align}
\nonumber (\mathbf{P1})~~~~~~~~~~~~~~~~ \min_{\mathbf{z}} ~~&\sum_n z_n
\\ \nonumber \text{Subject to} ~~~~&P_e(\mathbf{z}) - P^{DRML}_e \leq \epsilon,
\\				&z_n \in \mathcal{Z},
\label{eq:P1}
\end{align}
\vspace{-3mm}}

\noindent where $\mathbf{z}=[z_1,\cdots,z_R]$, and $\epsilon$ is a small positive value. Note that the objective function ensures that low-complexity detectors are selected, while the constraint ensures that the selected detectors are reliable.

\section{Proposed Method}
\label{sec:proposed_method}
The selection problem in ($\mathbf{P1}$) is a high-dimensional problem since $R$ is typically large and the analytical expression of $P_e(\mathbf{z})$ in terms of distributions of $\mathbf{y}_n, \mathbf{H}_n$ is extremely complex. Further, due to integer variables $z_n$, it is a non-convex, NP-hard problem. The communication receiver needs to quickly obtain a solution using a low-complexity method. Therefore, we decompose the problem into smaller sub-problems for each RE and select the detector such that the RE error rate with selected detector matches the RE error rate when DRML is used. The RE error event, denoted by $E_n(z_n)$, occurs while detector $z_n\in \mathcal{Z}$ is used and any one hard decision bit in the $n$th RE is in error, i.e., $b^{(z_n)}_{n,t,m} \neq b_{n,t,m}$ for any $t\in [1,2], m\in [1,M]$, where  $b^{(z_n)}_{n,t,m}$ is the hard decision bit defined as $b^{(z_n)}_{n,t,m} = 1, \text{ if } L^{(z_n)}_{n,t,m}>0,$ and $0$ otherwise.
\begin{comment}
{\small \begin{align}
\nonumber b^{(z_n)}_{n,t,m} & = 1, \text{ if } L^{(z_n)}_{n,t,m}>0,
\\b^{(z_n)}_{n,t,m} &= 0, \text{ otherwise}.
\label{eq:llr2bits}
\end{align}}
\end{comment}
The detector $z_n$ is selected independently for each RE by solving $(\mathbf{P2})$:

{\small 
\vspace{-3mm}
	\begin{align}
	\nonumber  (\mathbf{P2})~~~~~~~~~~ &\min_{z_n} ~~ z_n
	\\ \nonumber \text{Subject to}~~~ &\Pr(E_n(z_n)) = \Pr(E_n(5)),
	\\				&z_n \in \mathcal{Z},
\end{align}
\vspace{-3mm}}

\noindent where $E_n(5)$ is the RE error when DR-ML detector is selected. We argue that $z_n$s obtained by solving ($\mathbf{P2}$) essentially provide a feasible solution for ($\mathbf{P1}$). As mentioned before, one of the main challenges is to solve $(\mathbf{P2})$ in \textit{online detector selection} without actually evaluating (\ref{eq:LLR}) for any detector, i.e., the problem should be solved without running any candidate detector block. Further, the knowledge of transmitted bits $b_{n,t,m}$ is not available in \textit{online detector selection} and the analytical expression of $\Pr(E_n(z_n))$ is still a very complex function of distributions of $\mathbf{y}_n, \mathbf{H}_n$ and $\sigma^2$. Therefore, we train MLP network to obtain solution to ($\mathbf{P2}$) using features derived from $\mathbf{y}_n, \mathbf{H}_n, \sigma^2$. 

\begin{figure*}[t]
	\centering
	\includegraphics[width=1.5\columnwidth]{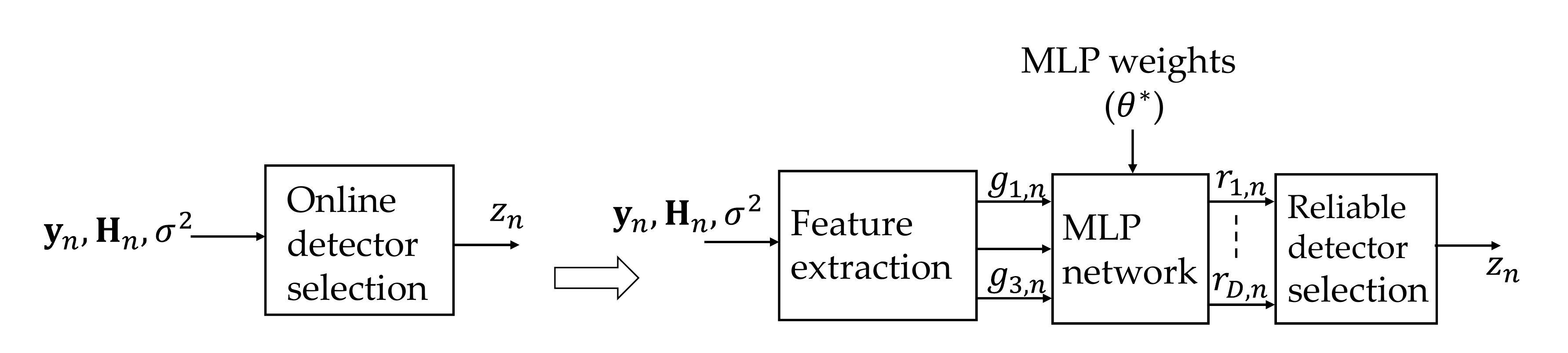}	
	\caption{\footnotesize Online detector selection.}
	\label{fig:detector_selection}
	\vspace{-6mm}
\end{figure*}

\begin{figure*}[t]
	\centering
	\includegraphics[width=1.5\columnwidth]{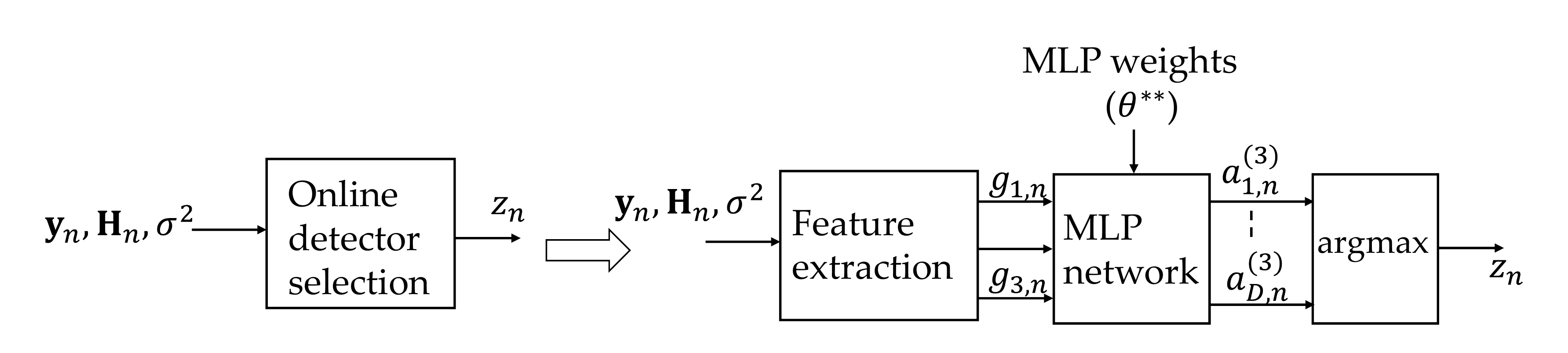}	
	\caption{\footnotesize Online detector selection after MLP re-training.}
	\label{fig:detector_selection_retraining}
	\vspace{-2mm}
\end{figure*}

First, a training dataset is generated offline that consists of features that are already available in communication receiver chain and the detector labels. The labels are essentially the solutions of $(\mathbf{P2})$ obtained using the genie knowledge of transmitted bits $b_{n,t,m}$. Next, the MLP is trained using quasi-Newton method to obtain network parameter $\theta^*$. As shown in Fig. \ref{fig:detector_selection}, we propose a \textit{reliable detector selection} method to further process MLP outputs to select the detector. Finally, the MLP is re-trained to incorporate \textit{reliable detector selection} into the MLP training itself and obtain network parameter $\theta^{**}$. Re-training reduces the complexity of online detector selection as shown in Fig. \ref{fig:detector_selection_retraining} and requires only \textit{argmax} operation after MLP in order to select the detector. Details of the proposed method are provided in the following sections.


\subsection{Generation of labeled dataset}
\label{sec:dataset}
The labeled dataset is generated offline and it consists of $N$ samples of tuple $\{g_1, g_2,\cdots, g_F,z\}$ containing features $\{g_1, g_2,\cdots, g_F\}$ and detector labels $z$ for each RE. 
\subsubsection{Label generation}
\label{sec:label_gen}
The labels $z_n$ are generated to provide a solution to ($\mathbf{P2}$). Let $E^c_n(5)$ be the event when RE error does not occur when DRML detector is used. If $E^c_n(5)$ occurs, then the events $E^c_n(z_n), z_n<5$ may also occur. In other words, if the DRML detector provides no RE error, then a lower-complexity detector $z_n<5 $ may be used if it also results in no RE error. However, if the event $E_n(5)$ occurs, then events $E^c_n(z_n), z_n<5$ will occur with very negligible probability. The event $E_n(5)$ occurs because noise sample $\mathbf{n}_n$ is large in that RE. We do not include such RE in MLP training as it introduces unnecessary noise in the training. Therefore, labels are generated using REs in which the event $E^c_n(5)$ occurs.

In offline label generation, each candidate detector $d\in \mathcal{Z}$ is evaluated for each RE to compute LLRs $L^{(d)}_{n,t,m}$ and hard decisions $L^{(d)}_{n,t,m}$. If the event $E^c_n(5)$ occurs, then the label is generated using genie information of transmitted bits $b_{n,t,m}$ as $z_n = \min \{d \given  b^{(d)}_{n,t,m} = b_{n,t,m}, d\in \mathcal{Z}, \forall t, \forall m\}$. Note that such labeling ensures that $\Pr(E_n(z_n))=\Pr(E_n(5))$ with the smallest $z_n$.

\subsubsection{Feature identification}
\label{sec:feature_id}
The input features are generated using $\mathbf{y}_n, \mathbf{H}_n, \sigma^2$, since LLRs are functions of these variables. We consider 7 candidate features which are functions of normalized signal and channel matrix:

{\small
\vspace{-3mm}	
\begin{align}
\mathbf{\bar{y}} = \frac{\mathbf{{y}}}{\sigma}, \mathbf{\bar{H}} = [\mathbf{\bar{h}}_{1}, \mathbf{\bar{h}}_{2}] = \frac{\mathbf{{H}}}{\sigma}
\end{align}
\vspace{-3mm}
}

\noindent The features are listed in Table \ref{table:features}. The variables $r_{ij,n}$ in the table are obtained from the QR factorization: 
{\small\begin{align}
\mathbf{\bar{H}}_n = \mathbf{Q}_n \mathbf{R}_n = \mathbf{Q}_n \left[\begin{matrix}
r_{11,n} &r_{12,n} \\ 
0 &r_{22,n}
\end{matrix}
\right]
\end{align}
\vspace{-3mm}}

\begin{table}
	\centering
	\caption{List of candidate features}
	\begin{tabular}{|c|c|c|c|}
		\hline
		& \textbf{Feature}&&\textbf{Feature} \\ &&& \\
		\hline
	$g_{1}$ & $|r_{22}|^2 $  &	$g_{5}$& Channel power: \\&&& ${||\mathbf{\bar{h}}_{1}||^2 + ||\mathbf{\bar{h}}_{2}||^2}$\\
	\hline
	&&& \\
	$g_{2}$ & y-h ratio: $\frac{|\mathbf{\bar{y}}^* \mathbf{\bar{h}}_{1}|^2}{|\mathbf{\bar{y}}^* \mathbf{\bar{h}}_{2}|^2}$ &$g_{6}$& $|r_{11}|^2$ \\
	\hline
	&&& \\
	$g_{3}$ & Min. eigenvalue of ${\mathbf{\bar{H}}^*\mathbf{\bar{H}}}$ &$g_{7}$& $|r_{12}|^2$\\
	\hline
	&&& \\
	$g_{4}$ & Max. eigenvalue of ${\mathbf{\bar{H}}^*\mathbf{\bar{H}}}$ &&\\
	\hline
\end{tabular}
\label{table:features}
\vspace{-4mm}
\end{table}	
We assume that these features are already computed in the communication receiver chain and require no additional computations. The labeled dataset of $N$ samples consisting of tuple of the candidate features $g_{1,n},\cdots,g_{7,n}$ and corresponding detector labels $z_n$ is generated under different channel models, i.e., EPA5, EVA30 and over a range of SNRs. 

\begin{figure}
	\centering
	\includegraphics[width=\columnwidth]{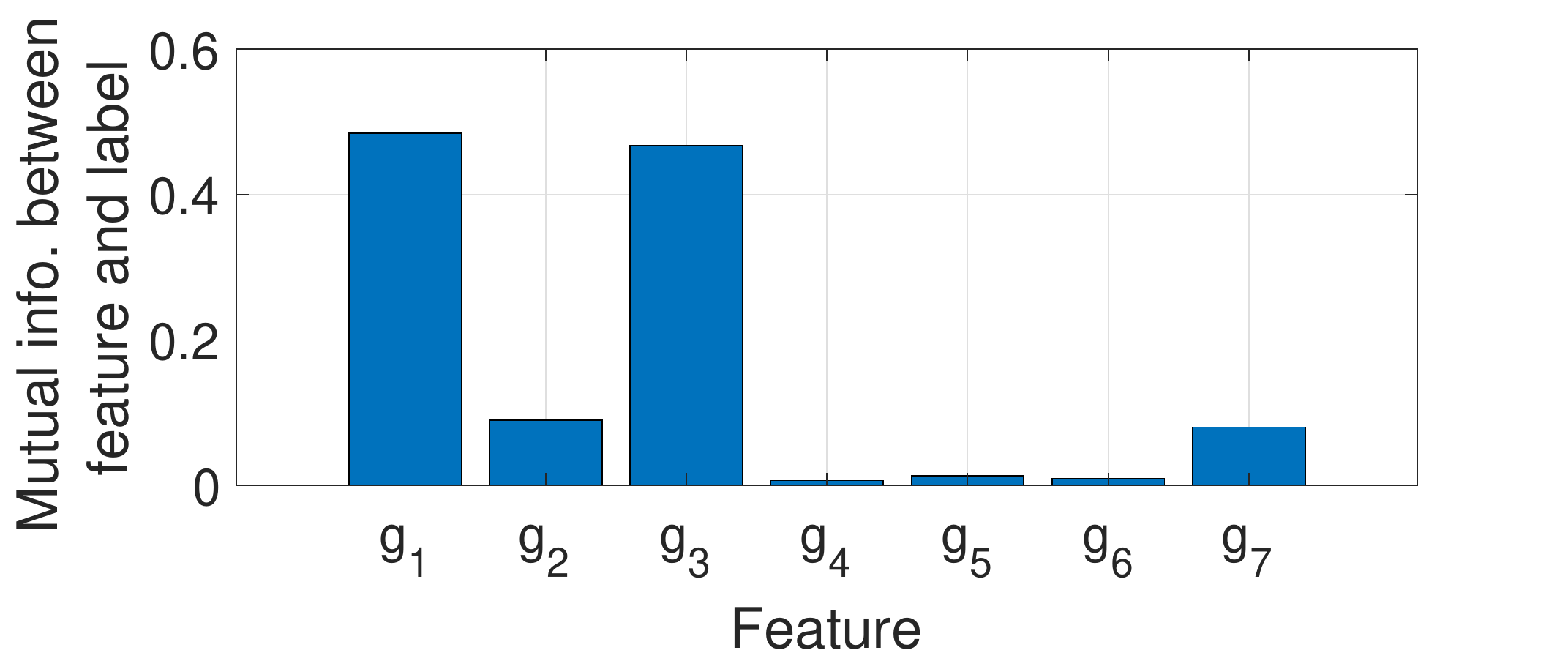}
	\caption{\footnotesize Mutual information $I(z;g_i)$ between feature and label.}
	\label{fig:mutual_info}
	\vspace{-6mm}
\end{figure}

Note that any other candidate features, that are informative about the detector label, may be considered. Intuitively, the features listed in the Table \ref{table:features} would be informative of the detector label. For example, if the minimum eigenvalue of $\mathbf{\bar{H}^*\bar{H}}$ is small, a high complexity detector would be required to get zero error in the RE since it means that symbol in one of the MIMO layers cannot be easily separated. Similar intuitive arguments can be made for other features. However, in order to mathematically identify relevant features for the MLP training, we rank the 7 candidate features in order of \textit{importance} using the mutual information based feature selection (MIFS) algorithm in \cite{battiti1994}.  The MIFS algorithm uses the mutual information $I(z;g_i)$ between labels $z$ and feature $g_i, i \in [1,7]$ and mutual information between features $g_i$ and $g_j$ to greedily search most relevant features. The algorithm provides feature ranking as: [$g_1$, $g_3$, $g_2$, $g_4$, $g_5$, $g_7$, $g_6$]. Therefore, feature $g_1$ is the most informative feature followed by feature $g_3$ and so on. We also observe that MI between feature $g_4$ and label  is very low as shown in Fig. \ref{fig:mutual_info}. Therefore, any feature ranked after $g_4$ does not provide any additional information about the label. The features ranked before $g_4$, i.e., features $g_1, g_2, g_3$ are selected for MLP training. Thus, the training dataset contains $N$ samples of tuple $\{g_{1,n}, g_{2,n}, g_{3,n}, z_n\}$. This training dataset is used to train the MLP network.

\subsection{MLP network}
\label{sec:MLP network}

We consider a 3-layer MLP network as shown in Fig.\ref{fig:mlp_full} with $3$ input features identified earlier, 1 hidden layer with $P$ nodes and output layer with $D$ outputs. In Fig.\ref{fig:mlp_full}, $w_{ij}^{(l)}$ is the weight between node-$j$ in layer-$l$ to node-$i$ in layer-$(l+1)$ and $v_{i}^{(l)}$ is the bias term between at node-$i$ in layer-$(l+1)$. The output of node-$i$ in layer-2 is given by
 
{\small
\vspace{-3mm}
\begin{align}
	a_{i,n}^{(2)} = f\left(\sum_{j=1}^{3}w_{ij}^{(1)} g_{j,n} + v_i^{(1)}\right), i\in [1,P],
	\label{eq:layer2_op}
\end{align}
\vspace{-3mm}}

\noindent where $f(x)=\frac{1}{1+e^{-x}}$ is the sigmoid activation function. Further, the output of node-$d$ in layer-$3$ is given by

{\small
\vspace{-3mm}	
\begin{align}
	a_{d,n}^{(3)} = \sum_{j=1}^{P}w_{dj}^{(2)} a_{j,n}^{(2)} + v_d^{(2)}, d\in [1,D].
	\label{eq:layer3_op}
\end{align}
\vspace{-4mm}}

\noindent Finally, the MLP output is given by applying \textit{softmax} as:

\begin{figure*} [t]
	\centering
	\includegraphics[width=1.3\columnwidth]{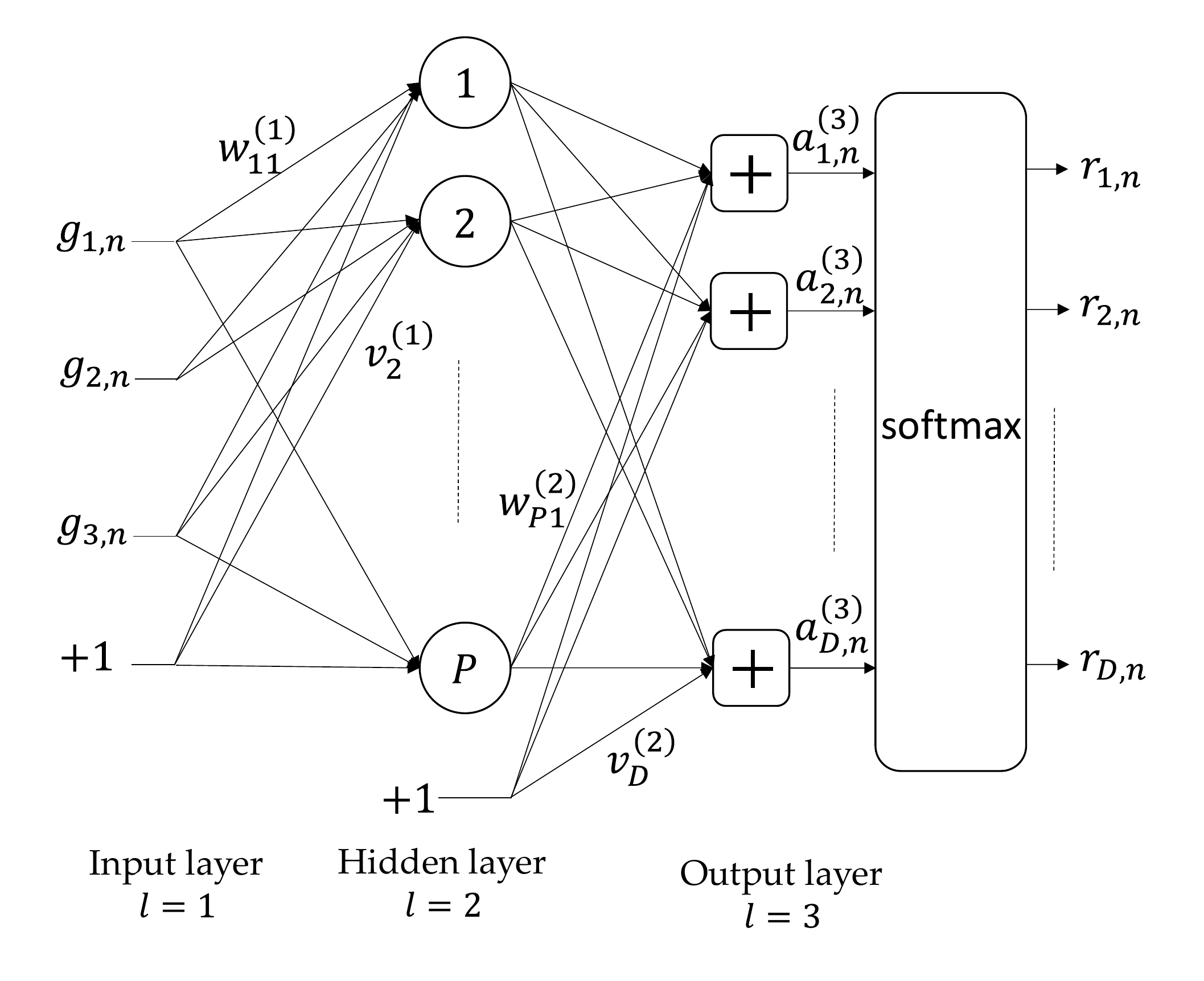}
	\caption{\footnotesize MLP network with $3$ input features, 1 hidden layer, $P$ hidden nodes, and $D$ outputs. The labels $+1$ indicate bias terms.}
	\label{fig:mlp_full}
	\vspace{-5mm}
\end{figure*}

{\small
\vspace{-4mm}
\begin{align}
	r_{d,n} = \frac{e^{a_{d,n}^{(3)}}}{\sum_{j=1}^{D}e^{a_{j,n}^{(3)}}}, d\in [1,D].
\end{align}
\vspace{-3mm}}

\noindent Note that softmax ensures that $0\leq r_{d,n}\leq 1$ and $\sum_d {r_{d,n}}=1$. The MLP network is trained offline using the training dataset $\{g_{n,1}, g_{n,2}, g_{n,3}, z_n\}$ in order to obtain  optimal weights and biases $w_{ij}^{(l)}, v_{i}^{(l)}$. The vector containing weights and biases is denoted by $\theta$. Thus the MLP network is parameterized by $\theta$. The optimal parameter $\theta^*$ minimizes the following cross-entropy cost function
\vspace{-3mm}
\begin{align}
c(\theta) = -\sum_{d=1}^{D} \sum_{n\in \mathcal{N}_d} \log(r_{d,n}),
\label{eq:cost}
\end{align}
\vspace{-2mm}

{\noindent}where $\mathcal{N}_d$ is the set of training samples of detector-$d$, i.e., samples with label $z_n=d$. The MLP is trained with quasi-Newton method to obtain the parameter $\theta^*$.

In order to improve the performance of the MLP training, we employ two pre-processing techniques on dataset, namely class re-sampling and class merging before MLP training.

\subsubsection{Class re-sampling and merging}
\label{sec:resampling}
The training dataset generally contains unequal number of samples for each detector class. We re-sample the training dataset to remove extra samples from detector class to keep maximum number of samples per class to $N_{max}=20,000$. Further, in static detector selection, it has been observed that the performance gap between detector-4 (ICR-64) and DR-ML is quiet small as seen in Fig. \ref{fig:BLER}, while the gap between detector-3 (ICR-16) and ML is higher. Therefore, in order to achieve performance similar to ICR-64, the MLP network must accurately learn which REs need to use ICR-64. However, we observed that the number of samples for class-4 is significantly smaller than number of samples for class-1,2,3. Therefore, the MLP cannot accurately predict which REs require ICR-64 detector. 

\begin{figure*}
	\centering
	\includegraphics[width=1.3\columnwidth]{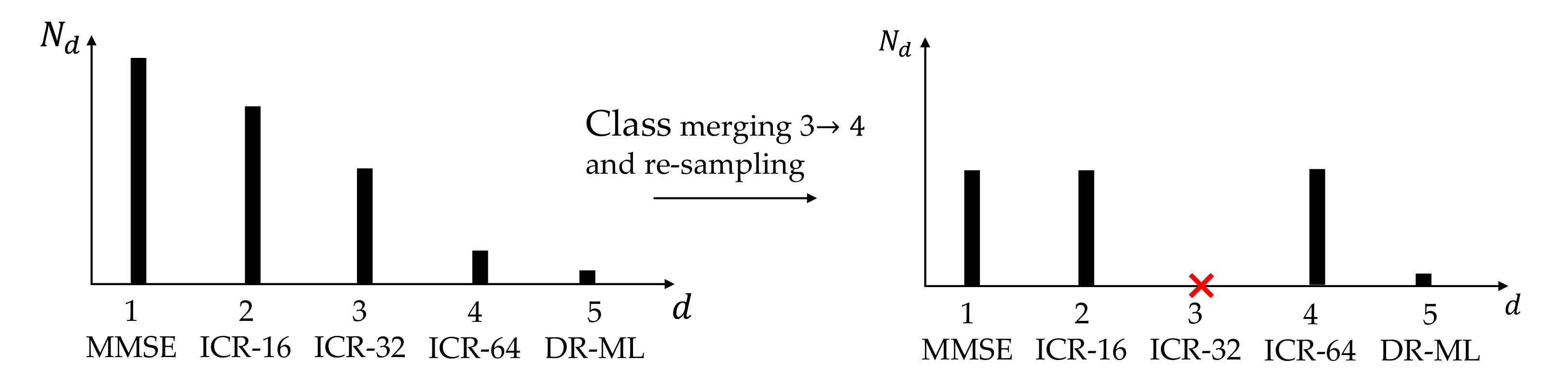}
	\caption{\footnotesize Class merging and re-sampling. $N_d$ is no. of samples in $\mathcal{N}_d$.}
	\label{fig:class_merging}
	\vspace{-5mm}
\end{figure*}

In order to overcome this issue, we merge class-3 with class-4 samples. Thus, samples with $z_n=3$ are converted to $z_n=4$. After class merging, the MLP is has sufficient number of samples for class-1,2 and 4 as depicted in Fig. \ref{fig:class_merging} and it can be trained to predict which REs require ICR-64 detector.

%\begin{comment}
%\end{comment}

\subsection{Reliable detector selection}
\label{sec:selection}
\begin{figure*}
	\centering
	\includegraphics[width=1.3\columnwidth]{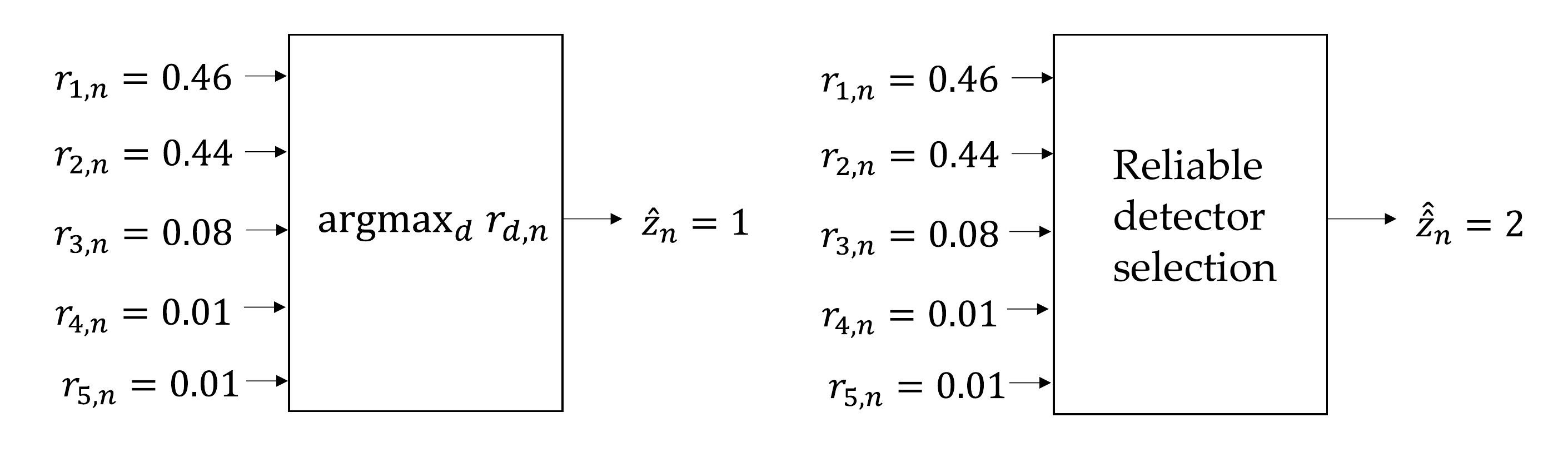}	
	\caption{\footnotesize Illustration of reliable selection assuming $\delta_1,\delta_2,\delta_3 = 0.1$.}
	\label{fig:reliable_selection}
	\vspace{-4mm}
\end{figure*}

After the MLP training, we obtain the parameter $\theta^*$ that minimizes the cost function in (\ref{eq:cost}). In online detector selection, the MLP network uses weights and biases in $\theta^*$ to provide outputs $r_{1,n},\cdots,r_{D,n}$. A naive way to select the detector is based on the maximum value among the MLP outputs as $\hat{z}_n = \arg\max_{d} r_{d,n}$.
\begin{comment}
\hat{z}_n = \arg\max_{d} r_{d,n}
\label{eq:armax}
\end{align}
\end{comment}
However, our objective is to select a reliable detector to generate LLR. The detector selected in this way may not be reliable, especially when the underestimation error occurs. It occurs when the detector label predicted by the MLP is lower than the true label in dataset. If there is an underestimation error, then a low complexity detector will be used for the RE that requires high complexity detector based on the channel conditions. This can increase the BLER after decoding. To limit the the underestimation error below a threshold $\gamma$, we use margin $\delta_d$ for selecting detector-$d$ instead of detector-$(d+1)$. A reliable detector is selected as
\begin{align}
\hat{\hat{z}}_n = \arg\min_{d\in [\hat{z}_n,D]} r_{d,n} - r_{d+1,n} > \delta_d.
\end{align}
An illustration of reliable detector selection is shown in Fig. \ref{fig:reliable_selection} where $\hat{z}_n=\arg\max_d r_{d,n}=1$, but $\hat{\hat{z}}_n=2$.

\begin{comment}
\begin{figure}
	\centering
	\begin{subfigure}[b]{0.5\textwidth}
		\includegraphics[width=\columnwidth]{Figures/reliable_selection_case1.pdf}	
		\caption{{\footnotesize Case-1.}}
		\label{fig:reliable_selection_case1}
	\end{subfigure}

	\begin{subfigure}[b]{0.5\textwidth}
		\includegraphics[width=\columnwidth]{Figures/reliable_selection_case2.pdf}
		\caption{{\footnotesize Case-2.}}
		\label{fig:reliable_selection_case2}
	\end{subfigure}
	\caption{\footnotesize Illustration of reliable selection. Assume margins for reliable selection $\delta_1,\delta_2,\delta_3 = 0.1$.}
	\label{fig:reliable_selection}
%	\vspace{-6mm}
\end{figure}
\end{comment}

\subsubsection{Computation of margins}
\label{sec:margins}
The margins $\delta_1, \delta_2,\cdots, \delta_{D-1}$ are computed offline after the MLP is trained and the parameter $\theta^*$ is computed. The features ${g_{1,n},g_{2,n},g_{3,n}}$ from the training dataset are fed to the MLP network and outputs $r_{1,n},\cdots, r_{D,n}$ are obtained for $n\in [1,N]$. For any given margin $\delta_d$, we can compute probability of correct estimation $\Pr(r_{d} - r_{d+1}>\delta_d | z=d)$ and probability of under-estimation $\Pr(r_{d} - r_{d+1}>\delta_d ; z=d+1)$ as follows:

{\small
\begin{align}
&\nonumber \Pr \left(r_{d} - r_{d+1}>\delta_d | z=d \right) = \frac{\sum \limits_n I(r_{d,n} -r_{d+1,n}> \delta_d ; z_n=d) }{\sum \limits_n I(z_n = d)},
\\&\nonumber \Pr \left(r_{d} - r_{d+1}>\delta_d ; z=d+1 \right)\\ &~~~~~~~~~~~~~~~~~~~~= \frac{\sum \limits_n I(r_{d,n} -r_{d+1,n}> \delta_d ; z_n=d+1) }{N},
\end{align}}
\noindent where $I(r_{d,n} -r_{d+1,n}> \delta_d ; z_n=d)$ is the indicator function ($=1$ when $n$-th sample satisfies $r_{d,n} -r_{d+1,n}> \delta_d$ and $z_n=d$). Then, the margins are set such that the probability of correct estimation is maximized while the under-estimation probability is kept below a threshold $\gamma$ by solving the following problem

{\small
\vspace{-3mm}
\begin{align}
\nonumber \max\limits_{\delta_d}~~~~& \Pr(r_{d} - r_{d+1}>\delta_d | z=d),
\\\text{subject to~~~} & \Pr(r_{d} - r_{d+1}>\delta_d ; z=d+1) < \gamma.
\label{eq:optimization}
\end{align}
\vspace{-3mm}}

\noindent We observe that $\Pr(r_{d} - r_{d+1}>\delta_d | z=d)$ and $\Pr(r_{d} - r_{d+1}>\delta_d ; z=d+1) $ are decreasing functions of $\delta_d$ and $0\leq r_{d,n}\leq 1$. Therefore, the solution to the above optimization problem is the smallest value $\delta_d$ between 0 and 1 that satisfies the constraint $\Pr(r_{d} - r_{d+1}>\delta_d ; z=d+1) < \gamma$. Each $\delta_d$ is found by sweeping the range from 0 to 1 and finding the smallest value that satisfies the constraint in (\ref{eq:optimization}). Note that $\gamma$ is a user specified threshold. If it is too high ($\gamma \rightarrow 1$), then the proposed method reduces to naive detector selection using $\argmax$ and it may increase BLER. On the other hand if $\gamma \rightarrow 0$, then high complexity detectors will be selected for most REs. 

\subsection{MLP re-training}
\label{sec:retraining}
In order to reduce the complexity of online detector selection, we incorporate \textit{reliable detector selection} in MLP re-training. For re-training, we generate new labels $\zeta_n$ for each RE using MLP parameter $\theta^*$ and margins $\delta_d$ as follows.  
The features $g_{1,n},g_{2,n},g_{3,n}$are fed at the input of the MLP network and outputs $r_{d,n}$ are computed using the parameter $\theta^*$. Then, the new labels are generated as
\begin{align}
\zeta_n = \arg\min_{d\in [z_n,D]} r_{d,n} - r_{d+1,n} > \delta_d.
\end{align}
The MLP is re-trained using the quasi-Newton method on the labeled dataset $\{g_{1,n},g_{2,n},g_{3,n}, \zeta_n\}$ to obtain new parameter $\theta^{**}$. In online detector selection, this new parameter is used as shown in Fig. \ref{fig:detector_selection_retraining}. Note that the \textit{softmax} and \textit{reliable detector selection} blocks are not required in online selection after MLP re-training. Further, the sigmoid function is replaced with piecewise linear approximation which has negligible complexity \cite{Tommiska2003}.

\vspace{-1mm}
\section{Results}
\label{sec:results}
\begin{figure*}[t]
	\centering
	\begin{subfigure}[b]{0.3\textwidth}
		\centering
		\includegraphics[width=\columnwidth]{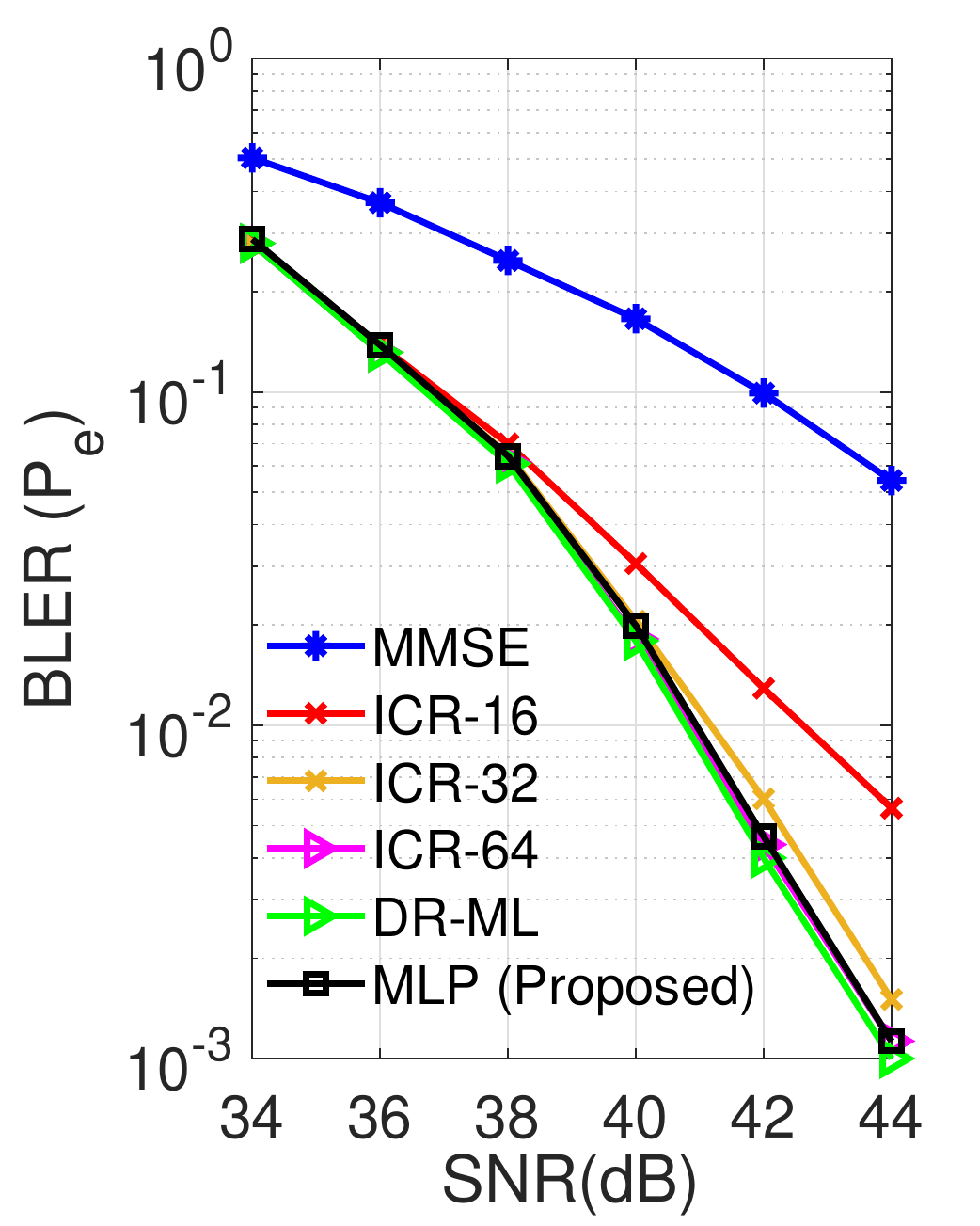}	
		\caption{{\footnotesize EPA-5.}}
		\label{fig:bler_EPA5}
	\end{subfigure}
	\begin{subfigure}[b]{0.3\textwidth}
		\centering
		\includegraphics[width=\columnwidth]{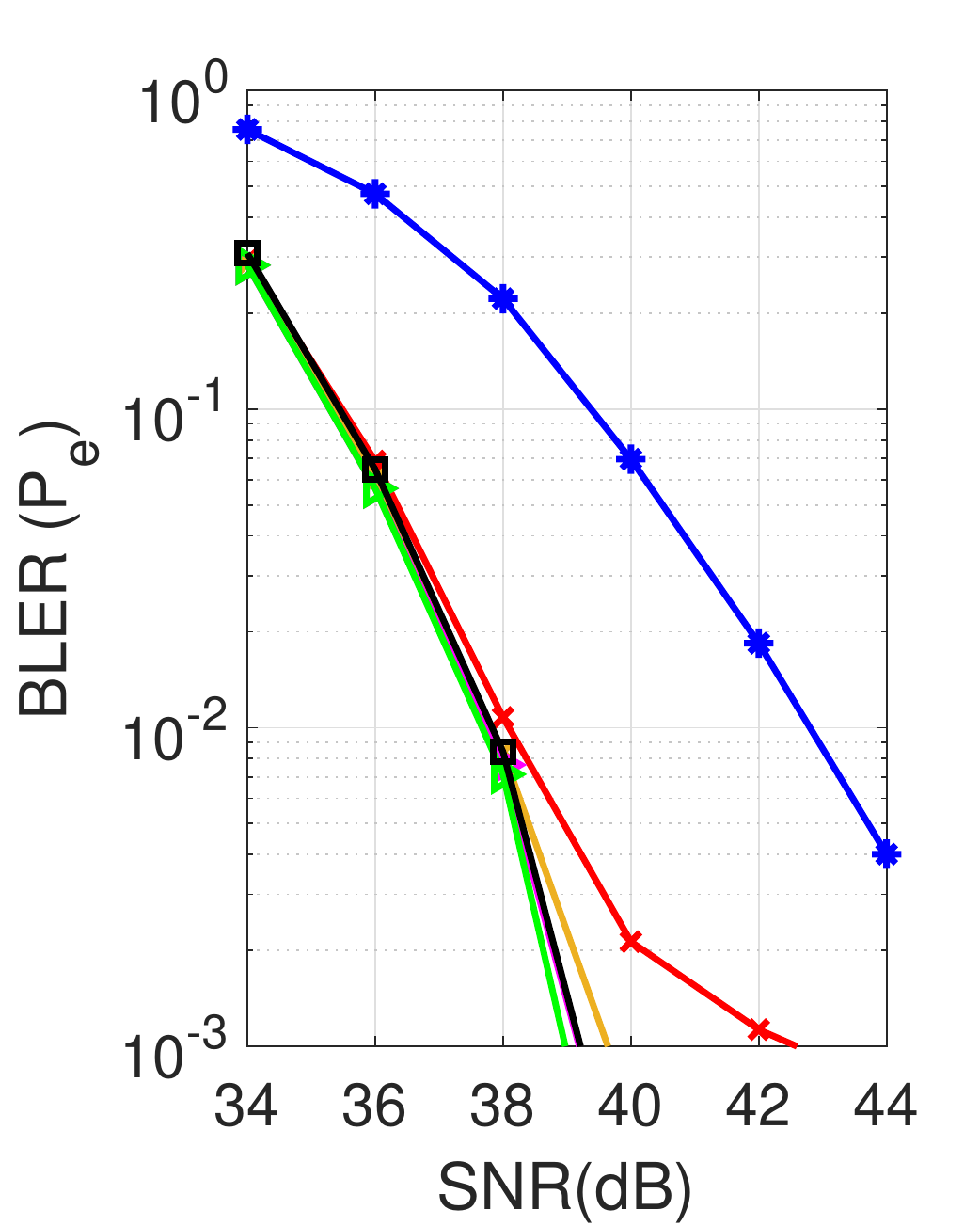}	
		\caption{{\footnotesize EVA-30}}
		\label{fig:bler_EVA30}
	\end{subfigure}
	\caption{\footnotesize BLER vs SNR for 256-QAM, MCS-27, coderate = 0.926}
	\label{fig:BLER}
	\vspace{-4mm}
\end{figure*}

\begin{figure*}[t]
	\centering
	\begin{subfigure}[b]{0.3\textwidth}
		\centering
		\includegraphics[width=\columnwidth]{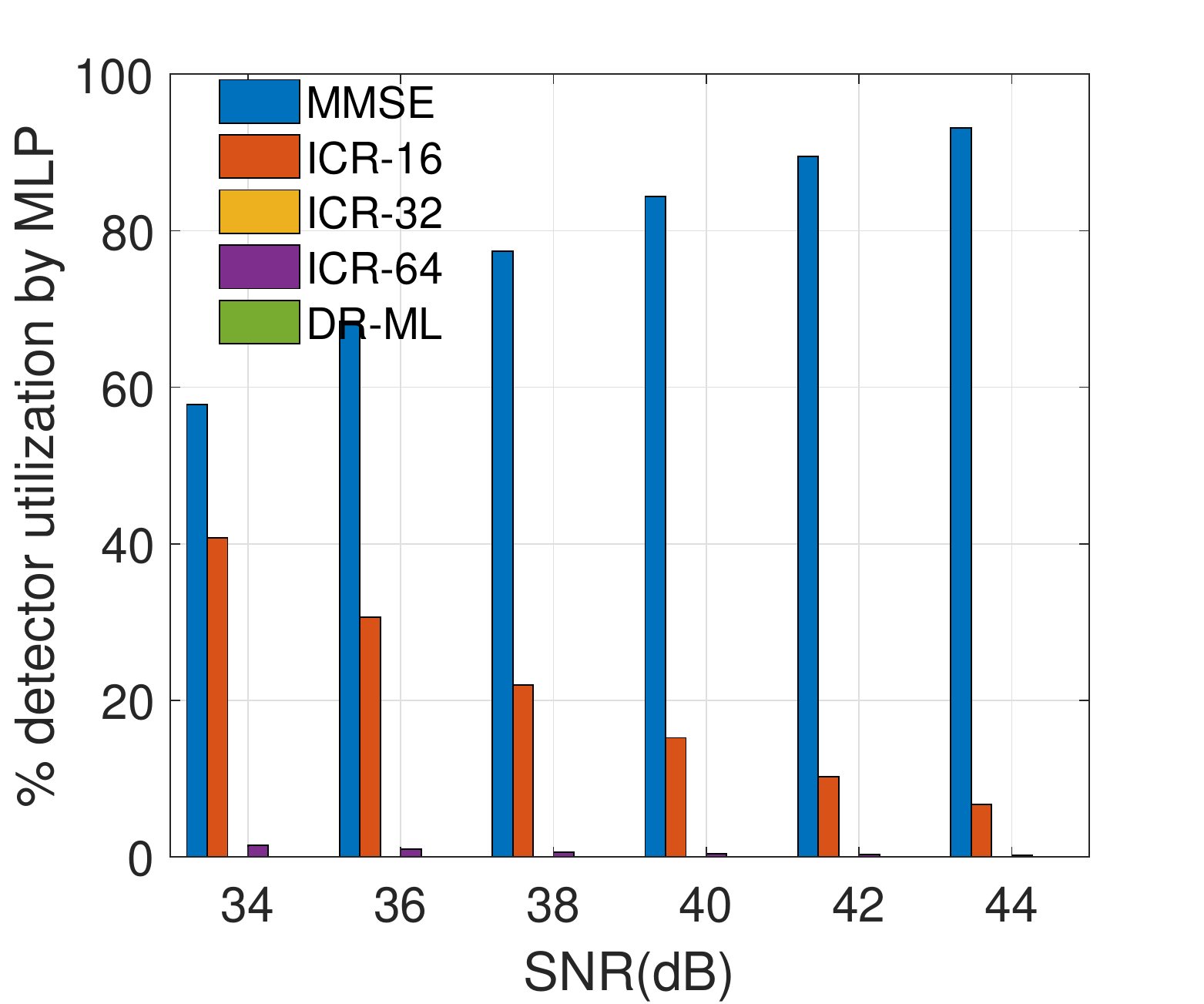}
		\caption{{\footnotesize EPA-5.}}
		\label{fig:utilization_EPA5}
	\end{subfigure}	
	\begin{subfigure}[b]{0.3\textwidth}
		\centering
		\includegraphics[width=\columnwidth]{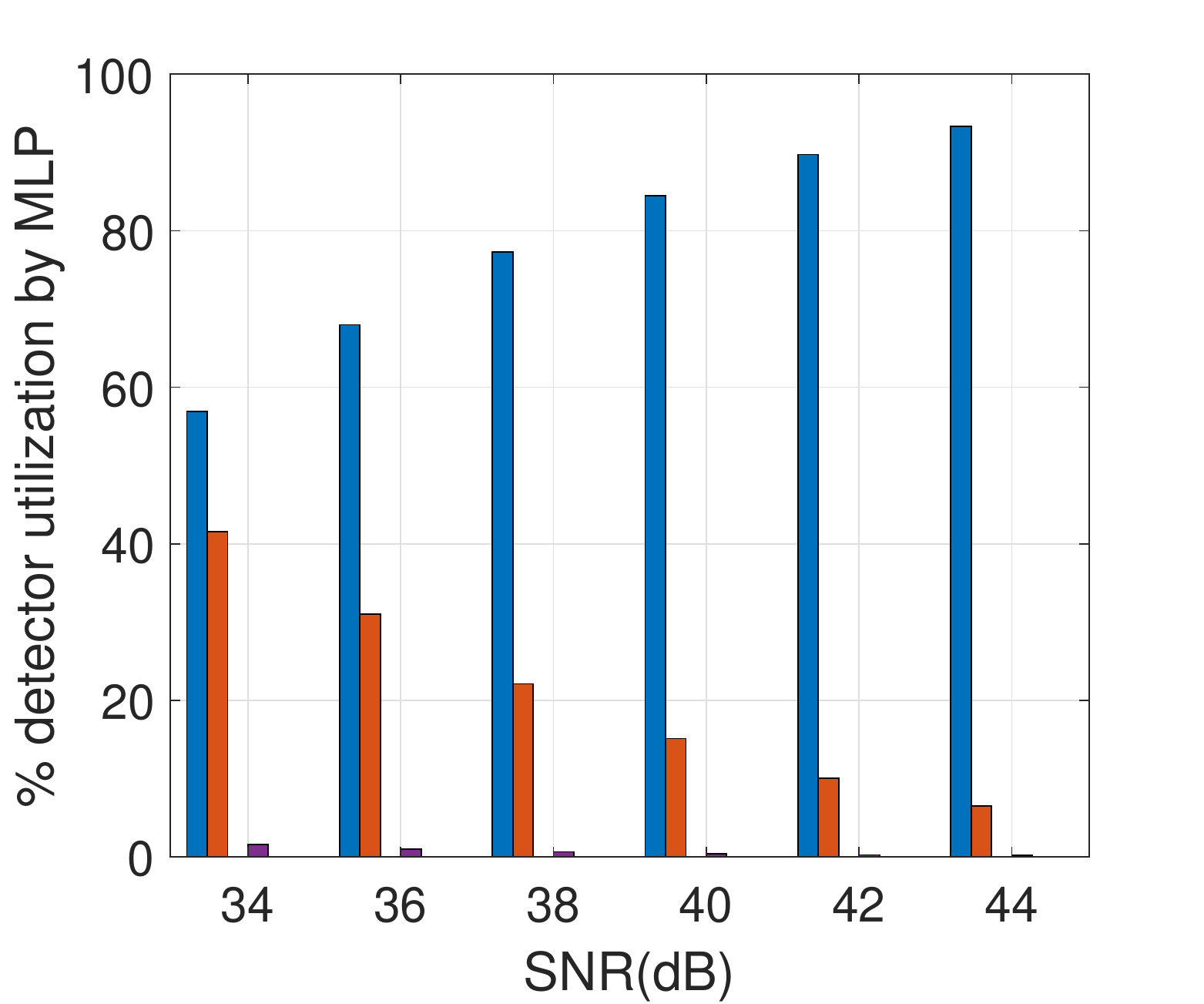}
		\caption{{\footnotesize EVA-30.}}
		\label{fig:utilization_EVA30}
	\end{subfigure}
	\caption{\footnotesize Detector utilization by proposed MLP based detector selection.}
	\label{fig:utilization}
	\vspace{-2mm}
\end{figure*}

The performance of the proposed method is evaluated using a MATLAB link level simulator that is developed to support 5G NR standard. We consider detector selection for 256-QAM modulation. Therefore, each MIMO layer in each RE carries $M=\log2(256)=8$ bits and the selected detector computes LLRs of 16 bits in the 2 MIMO layer in each RE. The transmitted OFDM signal has 15kHz subcarrier spacing, occupies 20MHz bandwidth and each of $8000$ transport block contains $R=7500$ REs. We use MCS-27 from the MCS table in \cite{3gpp2018_38214} with code-rate 0.926. First, a labeled dataset is generated offline by transmitting the signal under EPA5 and EVA30 channels and generating features and labels. After class-merging and re-sampling, an MLP network with $P=8$ nodes is trained to obtain $\theta^*$. Then, margins $\delta_d$ are computed for reliable detector selection using under-estimation threshold $\gamma=0.01$. Using the MLP parameter $\theta^*$ and margins $\delta_d$, the MLP is re-trained to obtain $\theta^{**}$. The BLER performance with the proposed online detector selection is shown in Fig. \ref{fig:BLER}. Utilization of various candidate detectors by the proposed detector selection is shown in Fig. \ref{fig:utilization}. Further, Table \ref{table:snr_gap_ed} shows the SNR gap between static detector selection and proposed dynamic detector selection along with average ED computations. We can see that the proposed method utilizes MMSE detector for most of the REs while achieving BLER similar to ICR-64 detector with significantly lower ED computations. Further, the BLER gap $({P_e(\mathbf{z}) - P^{DRML}_e})$ of $3.86 \times 10^{-4}$ is achieved at 40dB SNR under EPA5 and $5\times 10^{-4}$ is achieved at 38dB SNR under EVA30.

\subsubsection{Complexity analysis}
\label{sec:complexity_analysis}
The proposed method has significantly lower computational complexity as the number of ED computations is significantly lower as compared to any static detector selection as seen in Table \ref{table:snr_gap_ed}. For each RE, ED computation of $||\mathbf{y}_n - \mathbf{H}_n \mathbf{x}||^2$ requires 24 real multiplications and 21 real additions/subtractions. Similarly, the number of real additions/ subtractions and multiplications for the forward propagation of MLP are 77 and 64, respectively. We assume that the input features are computed using existing computational blocks in the modem and therefore, require no additional computations. The number of real computations per RE for ICR-64 and MLP method are shown in Table \ref{table:real_computations}. The proposed method provides similar BLER as ICR-64 (which is close to DR-ML), while reducing the number of multiplications by $12 \times$ under EPA5 and by $9.7\times$ under EVA30. The number of additions/subtractions are reduced by $10 \times$ under EPA5 and by $8.4\times$ under EVA30.

\begin{table*}[t]
  \centering
  \caption{SNR gap and ED computations per RE}
  \begin{tabular}{|c|c|c|c|c|}
  	\hline
  	{Detector selection} & \multicolumn{2}{c|} {SNR gap at 1\% BLER} &\multicolumn{2}{c|}  {Average ED computations}\\
  						 & \multicolumn{2}{c|} {from DR-ML(dB)}	  & \multicolumn{2}{c|}{per layer at 1\% BLER}\\
	\hline
  	              & EPA-5 & EVA-30 & EPA-5~~~~ & EVA-30 \\
  	\hline
%  	Static: MMSE  & $>3$   &$>4$   &$0$    &$0$\\
%  	\hline  	
%  	Static: ICR-16&$1.88$  &$0.41$     &$16$   &$16$ \\
%  	\hline  	
  	Static: ICR-32&$0.42$  & $0.14$      &$32$   &$32$ \\
  	\hline  	
  	Static: ICR-64&$0.09$   &$0.05$     &$64$  &$64$ \\  	
  	\hline
  	Static: DR-ML   &$0$ 	   &$0$    & $256$ & $256$\\  	
  	\hline
    \textbf{Dynamic: MLP} &$0.10$ &$0.05$ & $2.42$  & $4.81$ \\
%	\textbf{(Proposed)}   &&&& \\
  	\hline
  \end{tabular}
\label{table:snr_gap_ed}
\vspace{-4mm}
\end{table*}
\begin{table*}[t]
	\centering
	\caption{Computations per RE for 1\% BLER}
	\begin{tabular}{|c|c|c|}
		\hline
  	{Detector selection} & {Avg. no. of real} 		& {Avg. no. of real} 		\\
						 & {multiplications}	& {additions/subtractions}	\\
%		\hline		
%		Static: ICR-32   &$32\times 24 = 768$ 	   &$32 \times 21 = 672$   \\  	
		\hline		
		Static: ICR-64   &$64\times 24 = 1536$ 	   &$64 \times 21 = 1344$   \\  	
		\hline
		\textbf{Dynamic: MLP} &(under EPA5) $128.24$& (under EPA5) $133.21$\\
		\textbf{(Proposed)}   &(under EVA30) $158.55$ &(under EVA30) $159.73$ \\
		\hline
	\end{tabular}
	\label{table:real_computations}
	\vspace{-4mm}
\end{table*}

\section{Conclusion and Future Work}
\label{sec:Conclusion}
We proposed a low-complexity technique to select MIMO detector for each RE in the transport block using an MLP network. The MLP network is trained twice using features derived from received signal, channel matrix and noise variance to select a reliable detector for each RE. Results show that the proposed method can achieve BLER close to the case when dimension-reduced ML detector is used for all REs, while requiring significantly lower computational complexity in the detector block. Therefore, this technique can be used to significantly reduce power consumption in 5G NR modems. 

In future, this work can be extended to train one MLP network for different modulations, e.g., 64-, 256-, 1024-QAM over a large SNR range. Here, we considered detector selection for $2\times 2$ MIMO, but it can be easily extended for higher number of MIMO layers as well. 
%Further, the bit-widths required for MLP weights and input features can be optimized to adopt this concept in a real-world hardware.

%\renewcommand{\baselinestretch}{1.2}
\bibliographystyle{IEEEtran}
\bibliography{IEEEabrv,SBPreference}

\end{document}